\documentstyle[12pt]{article}
\renewcommand{\c}{\left(\frac{\pi}{18\,\lambda}\right)^{\frac{1}{6}}}
\newcommand{\cstl}{\mbox{cst}(\lambda)}
\newcommand{\pk}{\frac{\partial}{\partial k}}
\newcommand{\pkt}{\frac{\partial^2}{\partial k^2}}
\newcommand{\ak}{{\sqrt{k}\,\mbox{J}_{\frac{1}{3}}\left(\frac{2}{3}
\sqrt{\frac{\pi}{2\,\lambda}}\; |k|^{\frac{3}{2}}\right)}}
\newcommand{\ai}{\ak\,dk}

\newcommand{\Var}{\mbox{Var}}

\newcommand{\cor}{\mbox{Corr}}
\newcommand{\icor}{\mbox{ICorr}}
\newcommand{\iicor}{\mbox{Int}}
\newcommand{\n}{{N}}

\newcommand{\gaslager}{{\mbox{gas}}}
\newcommand{\ga}{\gamma}
\newcommand{\gas}{{\gaslager}}
\newcommand{\eq}{{\mbox{eq}}}
\newcommand{\heavi}{\theta}

\newcommand{\bt}{\beta}
\newcommand{\beg}{\begin{eqnarray}}
\newcommand{\beq}{\begin{eqnarray}}
\newcommand{\enq}{\end{eqnarray}}
\newcommand{\sig}{\sigma}
\newcommand{\cst}{{\rm constant}}

\newcommand{\ap}{\alpha}
\newcommand{\ta}{\tau}
\newcommand{\ro}{\rho}
\newcommand{\rb}{{\bar \rho}}
\newcommand{\pta}{{\frac{\partial}{\partial \tau}}}
\newcommand{\px}{{\frac{\partial}{\partial x}}}

\newcommand{\lam}{\lambda}
\newcommand{\Ai}{{\rm Ai}}
\newcommand{\Bi}{{\rm Bi}}
\newcommand{\Ga}{\Gamma}
\newcommand{\Co}{{\cor}}
\newcommand{\Ic}{{\icor}}
\newcommand{\IIc}{{\iicor}}
\newcommand{\Fl}{J_{1/3}\left(\frac{2}{3}\left(\frac{\pi}{2\lam}\right)^{1/2}
k^{3/2}\right)}
\newcommand{\cq}{{\cal Q}}
\newcommand{\vr}{{\rm Var}}
\newcommand{\cj}{{\cal J}}
\begin{document}

\title{Universal Parametric Correlations of Eigenvalues of
       Random Matrix Ensembles}
\author{Kasper Juel Eriksen$^{\dag}$ and Yang Chen\\
        Department of Mathematics, Imperial College\\
        180 Queen's Gate, London SW7 2BZ, U K\\
$^{\dag}$ {\O}rsted Laboratory,
        Niels Bohr Institute\\
        H.C.{\O}. Universitetsparken 5,
        2100 Kbh. {\O}, Denmark }
\date{\today}
\maketitle

\begin{abstract}
Eigenvalue correlations of random matrix ensembles as a function of an
external (parametric) perturbation are investigated via the
Dyson Brownian Motion model in the situation
where the level density has a hard edge singularity.
By solving a linearized hydrodynamical equation, a universal
dependence of the density-density correlator on the external field is found.
As an application we obtain a formula for the variance of linear
statistics with the parametric dependence exhibited as a Laplace transform.
\end{abstract}

\newpage

\noindent
{\bf 1 Introduction.}
\par\noindent
Eigenvalues of $N\times N$ matrices can be viewed as
energy levels $(x_a,\;\;a=1,\cdots,N)$ of an effective Hamiltonian
\beg
H_0=\sum_ax_an_a,
\enq
where $n_a$ are the occupation at level $a.$ A proposal by
Wigner\cite{Mehta} is that the levels are drawn from an
ensemble of matrices, and when
restricted to the eigenvalues has the joint probability distribution,
\beg
P(x_1,\cdots,x_N)\prod_{a=1}^{N}dx_a=C_N\;\;{\rm e}^{-\bt W(x_1,\cdots,x_N;u)}
\prod_{a=1}^{N}dx_a,
\label{jointprop2}
\enq
where
\beq
W(x_1,\ldots,x_\n;u) =-\,\sum_{a<b}\ln|x_a-x_b|+\sum_a u(x_a),
\enq
$C_N$ is a constant and
$\bt=2,1,4$ describes ensembles with unitary, orthogonal and symplectic
symmetries. The density of levels is defined by
\beg
\sig(x):=\left<\sum_a\delta(x-x_a)\right>_\eq,
\enq
where $<\cdots>_\eq$ is an average with weight $P$ of Eq. (2). For
$u(x)=x^2$ and $x_a$ supported on the real line, $\sig$
is given by the Wigner semi-circular law\cite{Mehta} in the limit of
large $N$,
\beg
\sig_{{\rm W}}(x)\sim {\sqrt {N-x^2}},\;\;x\in (-{\sqrt N},{\sqrt N}).
\enq
Other ensembles arising from transport in
disordered electronic systems\cite{Stone}, have $x_a$
supported on the right half line with
different confining potentials, $u(x)$. For $u(x)=x^{\ap}$ and
$\ap>1/2$
one finds the eigenvalue density\cite{linstat, Eriksen-Chen};
\beg
\sig(x)\sim {\sqrt {\frac{N}{x}}},\;\;x\ll N
\enq
is universal near the ``hard edge''\cite{Tracy}, in the sense that
it is independent of $\alpha$. It can be shown
that for $0<\alpha<1/2,$ $\sig(x)\sim 1/x^{1-\alpha}$\cite{Eriksen-Chen}.
In this paper we shall be interested in the response of the levels when
$H_0$ is perturbed by an external potential.
Of particular interest is
the eigenvalue correlator as a function of the external potential.
This problem
was first studied in the context of the energy eigenvalues distribution
of a disordered metallic ring subjected to an external magnetic field using
diagrammatic techniques\cite{Safer},\footnote{More precisely, the density of
state-density of state correlation function at different fluxes.} and
it was found that the eigenvalue correlations are universal
after an appropriate rescaling. These results
were later reproduced in\cite{Beenakker1} using the Brownian motion model
of Dyson\cite{Dyson1} in the hydrodynamical approximation. Exact correlations
for all strengths of the perturbation were obtained in\cite{Ben}
using the method of supersymmetry pioneered by Efetov\cite{Efetov}.
All of the above results are
valid in the bulk of the spectrum where the density is uniform;
$\sig_{{\rm bulk}}=\sig_{{\rm W}}(0)\sim \cst.$\\
The phenomenological theory proposed by Dyson\cite{Dyson1}
interprets the eigenvalues $x_a$ as
positions of classical particles which are governed by an
over-damped Langevin equation subjected to Gaussian random force $f_a(\ta)$,
\beg
\gamma \frac{dx_a}{d\ta}=-\frac{\partial W}{\partial x_a}
+f_a(\ta),
\enq
where $\gamma$ is the friction coefficient,
\beg
{\overline {f_a(\tau)}}=0,\;\;\;\;{\overline {f_a(\ta)f_b(\ta^{\prime})}}
=\frac{2\ga}{\bt}\delta_{ab}\delta(\ta-\ta^{\prime}),
\enq
and
$\tau$ is related to the strength of the perturbation, $X.$
Since the $x_a$'s undergo a Brownian motion it is to be expected that
$X^2\propto \tau$ \cite{Dyson1,Beenakkerbrown}. A Fokker-Planck equation that
describes the ``time'' dependent joint probability distribution can be
derived, and reads
\beg
\gamma\pta P(x_1,\cdots,x_N,\ta)=\sum_{a}\frac{\partial}{\partial x_a}
\left[\frac{\partial W}{\partial x_a} +
\bt^{-1}\frac{\partial}{\partial x_a}\right]P(x_1,\cdots,x_N,\ta),\;\;\ta>0,
\enq
subjected to the initial condition;
$$P(x_1,\cdots, x_N,0)=\prod_{a=1}^{\n} \delta(x_a - x_{a}^{0}),$$
where $x_{a}^{0}$ is the initial position of particle
$a$.
The stationary solution of the Fokker-Planck equation is
\beg
P(x_1,\cdots,x_N,\infty)=C_N\;{\rm e}^{-\bt W}.
\enq
It was shown by Dyson\cite{Dyson1} that the time dependent density,
\beg
\sig(x,\ta)=\left<\sum_a\delta(x-x_a(\ta))\right>_{\tau},
\enq
[here $<\cdots>_{\ta}$ denotes average with ``time-dependent'' weight
given in Eq.(9)] satisfies a non-linear conservation law in the
``hydrodynamical'' approximation.
The non-equilibrium density $\sig(x,\ta)$ evolves in $\ta$ according to
\beg
\pta\sig(x,\ta)=\px\left(\sig(x,\ta)\px \Psi\right),
\label{13}
\enq
where
\beg
\Psi(x,\ta)=u(x)-\int_{K}dy\sig(y,\ta)\ln|x-y|+
\left(\frac{1}{\bt}-\frac{1}{2}\right)\ln [\sig(x,\ta)].
\label{14}
\enq
Note that the ``time'' dependent density is normed to $N$;
$\int_{K}dx\sig(x,\ta)=N,$ where $K$ is the interval on
which the levels are supported.\\
The solution to Eq.(\ref{13}) with Eq.(\ref{14}), will enable us to
determine the parametric $(\tau)$ dependence  of quantities
related to the eigenvalues $x_a.$
\par\noindent
{\bf 2 Linearization.}\par\noindent
The stationary density, $\sig(x):=\sig(x,\infty)$,
of the non-linear diffusion equation satisfies a
self-consistent H\"uckel type equation,
\beg
u(x)-\int_{K}dy\sig(y)\ln|x-y|+\left(\frac{1}{\bt}-\frac{1}{2}\right)
\ln[\sig(x)]
=A=\cst.
\label{huckel}
\enq
This suggests that for sufficiently long time\footnote{More precisely
$D\ta/\ga L\gg 1,$ where $L$ is the interval over which the density
extends\cite{Dyson1}.},
we split the non-equilibrium density into an equilibrium part
plus a small perturbation, $\ro(x,\ta);$
\beg
\sig(x,\ta)=\sig(x)+\ro(x,\ta),
\label{linearsigma}
\enq
where $\sig(x)$ the equilibrium density. Substituting
Eq.(\ref{linearsigma}) into
Eq.(\ref{13}) and discarding all terms of $O(\ro^2),$ gives
\beg
\pta\ro(x,\ta)=-\px J(x,\ta),
\enq
where the ``particle'' flux is
\beg
J(x,\ta):=\frac{\sig(x)}{\ga}\px\int_{K}\ro(y,\ta)\ln|x-y|dy.
\enq
This is unlike the ordinary diffusion equation in that the particle flux
requires the entire distribution, $\ro$, to specify its value at one point.
As an example, we consider the Gaussian ensembles with $u(x)=x^2$,
and $K=(-{\sqrt N},{\sqrt N}).$ In the
$N\rightarrow \infty$ limit, and scaling into the bulk $x\ll N$, where
the density is uniform, $\sig(x)=D={\rm constant}$, the diffusion equation
becomes,
\beg
\pta\ro(x,\ta)=-\frac{D}{\ga}\frac{\partial^2}{\partial x^2}
\int_{-\infty}^{+\infty}dy\ro(y,\ta)\ln|x-y|.
\enq
This is converted into
\beg
\pta{\rb}_k(\ta)=-\frac{\pi D}{\ga}|k|{\rb}_k(\ta),
\enq
via a Fourier transform\cite{Dyson1,Beenakker1},
${\rb}_k(\ta)=\int_{-\infty}^{+\infty}dx{\rm e}^{ikx}\ro(x,\ta),$ with the
solution;
\beg
{\rb}_k(\ta)={\rb}_k(\ta^{\prime})
{\rm e}^{-\frac{\pi D}{\ga}|k|(\ta-\ta^{\prime})},
\;\;\;\ta\geq \ta^{\prime}.
\enq
{}From the  $\ta$ dependence of ${\rb}_k$, we can infer the
$\ta$ dependence of
\beq
{\rm Corr}(x,y,\ta)
& := &
\left<\sig(x,\ta)\sig(y,0)\right>_\eq-\left<\sig(x,\ta)\right>_\eq
\left<\sig(y,0)\right>_\eq\\
&=&
\int_{-\infty}^{+\infty}\frac{dk}{2\pi}
\int_{-\infty}^{+\infty}\frac{dp}{2\pi}
{\rm e}^{-ikx-ipy-\frac{\pi D}{\ga}|k|\ta}{\rm Corr}(p,k),
\enq
where $$<\phi>_\eq:=\left(\prod_{a}\int_{-\infty}^{+\infty}dx^0_a\right)
P(x^0_1,\cdots,x^0_N)\phi(x^0_1,\cdots,x^0_N)$$
denotes an average over the initial condition.
Here ${\rm Corr}(p,k)$ is the Fourier transform of the
equilibrium density-density correlation function\cite{Mehta};
${\rm Corr}(x-y)=-\frac{1}{\pi^2\bt}\frac{1}{(x-y)^2},\;x\neq y.$
\footnote{The exact density-density correlation function for
the Gaussian ensemble ($\bt=2$, and $x\neq y$) is
$-\left[\frac{\sin\pi(x-y)}{\pi(x-y)}\right]^2$\cite{Mehta}. The
hydrodynamical approximation essentially replaces
$[\sin\pi(x-y)]^2$ by 1/2.}
\beg
{\rm Corr}(p,k)=\int_{-\infty}^{+\infty}dx\int_{-\infty}^{+\infty}
dy{\rm e}^{ikx+ipy}{\rm Corr}(x-y)=\frac{2}{\pi\bt}\delta(k+p)|k|.
\enq
Therefore, $${\rm Corr}(x,y,\ta)=\frac{1}{\bt\pi^2}
\Re \frac{1}{\left[\frac{D\pi\ta}{\ga}+i(x-y)\right]^2}$$
is the universal ``time'' dependent density-density correlation function
when $\ta$ is measured in units of $\frac{\ga}{\pi D}$
\cite{Safer,Beenakker1,Ben}\footnote{\cite{Ben} gives the exact result
for all $\tau\geq 0$ and $\bt=2,\;1,\;4.$}
\\
\noindent
{\bf 3 Hard edge correlations.}\\
We shall now focus our attention on ensembles where the eigenvalues are
supported on the right half-line and
$\sig(x)$ has the universal square root singularity at the origin.\\
${\rm Corr}(x,y,\ta)$ and ${\rm Int}(x,y,\ta)---$the twice integrated version
of ${\rm Corr}(x,y,\ta)$ are derived in Section {\bf 3}.
These are applied to determine the parametric
dependence of the variance of arbitrary linear statistics, given by
Eqs. (51) and (54).
As an application, we compute in Section {\bf 4}
the variance of $\sum_a\left[1+x_a(\tau)\right]^{-1},$
which gives the conductance fluctuation of a quasi-one dimensional
disordered system as a function of the external perturbation. According to
the Landauer formula, the conductance, $g$, is $\sum_{a}[1+x_a]^{-1}.$
\cite{Stone} \\
We have therefore solved in the hydrodynamical approximation
the problem posed in\cite{Beenakker1} for the case where the equilibrium
eigenvalue density displays a hard edge singularity;
$\sig(x)\sim D/{\sqrt x}.$ Since translational
invariance is no longer valid, results obtained in the bulk scaling
limit of the Gaussian ensembles (with $u(x)=x^2$) are no longer applicable
here\cite{Macedo}. For example, the gap formation probability,
$E_{\bt}(0,s)$, of the Laguerre ensembles
(with $u(x)=x-\alpha\ln x,\;x>0,\;\alpha>-1.$)
is distinct from that of the Gaussian ensembles\cite{Tracy,Chen,GFP}.
We expect ${\rm Corr}(x,y,\ta)$ to have
distinct ``time'' decay modes from that found in the bulk scaling case
\footnote{An application of a
dimensional argument on Eq. (18), obtained by scaling into the bulk of the
spectrum, shows that the typical distance
covered by a diffusing particle over time $t$ is $|x|\sim t,$ which is
``faster'' then Einstein diffusion; $|x|\sim {\sqrt t}.$
Looking ahead to Eq. (24)
the same analysis shows that $|x|\sim t^{2/3}.$ This suggests that
particle transport near the hard edge is intermediate between ballistic
motion and classical diffusion.}. \\
\noindent
The diffusion equation now reads
\beq
\frac{\partial \,\ro(x,t)}{\partial t}
=-\px\left(\frac{1}{{\sqrt x}}{\rm P}
\int_{0}^{\infty}dy\frac{\ro(y,t)}{x-y}\right),
\label{difffortime}
\enq
where $D\ta/\ga=t.$ [We have written $\rho(x,\tau)=\rho(x,t).$]
With the ansatz $\ro(x,t)\propto \ro(x,\lam){\rm e}^{-\lam t}$, Eq.(24)
becomes,
\beg
\lam\ro(x,\lam)=\px\left(\frac{1}{{\sqrt x}}\int_{0}^{\infty}
dy\frac{\ro(y,\lam)}{x-y}\right),\;\;\;\lam>0.
\label{diffformodes}
\enq
The boundary condition on $\rho(x,\lambda)$
is such that the particle flux vanishes at the boundaries;
$\lim_{x\rightarrow 0}J(x,\lam)=0,$ and
$\lim_{x\rightarrow \infty}J(x,\lam)=0.$ The boundary condition at $x=0$ reads,
\begin{equation}
\lim_{x \rightarrow 0} \frac{1}{\sqrt{x}}{\rm P}
\int_{0}^{\infty}\frac{\rho(y,\lambda)}{x-y}\,dy=0.
\label{fluxboundarycondition}
\end{equation}
\renewcommand{\ro}{\tilde{\rho}}
\noindent
Since ${\rm P}\int_{0}^{\infty}\frac{1}{{\sqrt y}(x-y)}dy=0,\;\;x>0,$
we find it convenient to write $\rho(x,\lambda)$ as
\begin{equation}
\rho(x,\lambda) = \ro(x,\lambda) + \frac{\cstl}{\sqrt{x}},
\label{expressforrho}
\end{equation}
where $\ro(x,\lambda)$ fulfills a stronger condition,
\begin{eqnarray}
\int_{0}^{\infty}dy\frac{\ro(y,\lam)}{y}=0,
\label{boundarycononrho}
\end{eqnarray}
and ${\rm cst}(\lam)$ to be determined later is a function of
$\lambda$ only. In appendix \ref{appchechsol} it is shown that the final
solution (Eq. (42)) fulfills the boundary condition
(\ref{fluxboundarycondition}).
The term $\frac{\cstl}{\sqrt{x}}$ is similar to
the equilibrium solution and we conjecture that the
solution for a general hard edge density always
has this structure. We now go on to solve
Eq. (\ref{diffformodes}). This can be accomplished by
``un-folding'' the half-line into the
real-line\cite{Akhiezer}. With the change of variables, $v={\sqrt y},$
$u={\sqrt x},$ we have
\beg
\int_{0}^{\infty}dy\frac{\rho(y,\lam)}{x-y}
&=&
\int_{0}^{\infty}dy\frac{\ro(y,\lam)}{x-y}
=\int_{0}^{\infty}dv\;\frac{2v\ro(v^2,\lam)}{u^2-v^2}
\\
&=&\int_{0}^{\infty}dv\frac{\ro(v^2,\lam)}{u-v}+\int_{0}^{\infty}dv\;
\frac{-\ro(v^2,\lam)}{u+v}.
\enq
Now introduce an odd function of $v$,
\beg
\ro_1(v,\lam)&\equiv& \ro(v^2,\lam),\;\;\;\;v>0
\\
\ro_1(v,\lam)&\equiv &-\ro(v^2,\lam),\;\;\;v<0
\enq
and find
$$
\int_{0}^{\infty}dy\frac{\rho(y,\lam)}{x-y}
=\int_{0}^{\infty}dv\frac{\ro_1(v,\lam)}{u-v}+\int_{0}^{\infty}
dv\frac{\ro_1(-v,\lam)}{u+v}$$
\beg
=\int_{0}^{\infty}dv\frac{\ro_1(v,\lam)}{u-v}+\int_{-\infty}^{0}
dv\frac{\ro_1(v,t)}{u-v}
=\int_{-\infty}^{+\infty}dv\frac{\ro_1(v,\lam)}{u-v}.
\label{rewritero1}
\enq
With Eq. (\ref{rewritero1}) and Eq. (\ref{expressforrho}) the diffusion
equation, Eq. (\ref{diffformodes}), becomes,
\beg
\frac{\cstl}{u} + \lam\ro_1(u,\lam)=
\frac{1}{2u}\frac{\partial}{\partial u}
\left(\frac{1}{u}\int_{-\infty}^{+\infty}dv\frac{\ro_1(v,\lam)}{u-v}\right).
\label{diffforro1}
\enq
Although Eq. (\ref{diffforro1}) is derived for $u>0$ it is also valid for
$u<0.$ This can be seen from the fact that $\ro_1$ is an odd function of $u.$
To proceed further, we introduce the even function,
\beg
\ro_2(u,\lam):=\frac{\ro_1(u,\lam)}{u},
\enq
and with its aid we find $\ro_2(u,\lam)$ satisfies,
\beg
2\,\lambda\,\cstl + 2\,\lambda\,u^2\,\ro_2(u,\lam)
=\frac{\partial}{\partial u}\int_{-\infty}^{+\infty}
dv\frac{\ro_2(v,\lam)}{u-v}
\label{diffforro2}.\enq
We have made use of the flux condition at origin, Eq.
(\ref{boundarycononrho}),
\beg
\int_{-\infty}^{+\infty}du\ro_2(u,\lam)=2\int_{0}^{\infty}du\ro_2(u,\lam)
=0,
\label{boundarycononro2}
\enq
to arrive at Eq. (\ref{diffforro2}).
This accomplishes the un-folding of half-line diffusion equation
onto the real line. Clearly, Eq. (\ref{diffforro2}) can be be
solved by a Fourier transformation.
\\
Eq. (\ref{boundarycononro2}) implies that $\ro_2(u,\lam)$ is an
oscillatory function of $u$ and its Fourier transform,
\beg
\rho_2(k,\lam):=\int_{-\infty}^{+\infty}du{\rm e}^{iku}\ro_2(u,\lam),
\enq
vanishes at $k=0.$ A further condition is that $\rho_2$ is even in $k.$\\
Once $\rho_2(k,\lam)$ is found, the original density can be
recovered by standard inversion formulas;
\beg
\frac{\ro(x,\lam)}{{\sqrt x}}=\frac{1}{\pi}\int_{0}^{\infty}
dk\cos\left(k{\sqrt x}\right)\rho_2(k,\lam).
\label{recoverro}
\enq
A simple calculation shows that the transformed density, $\rho_2(k,\lam)$,
satisfies an Airy equation with a point source,
\beg
\frac{d^2}{dk^2}\rho_2(k,\lam)+\frac{\pi|k|}{2\lam}\rho_2(k,\lam)
&=&
2\,\pi\,\cstl\,\delta(k).
\enq
The solution of this is a linear combination of
$\Ai\left(-(\pi/2\lam)^{1/3}|k|\right)$ and
$\Bi\left(-(\pi/2\lam)^{1/3}|k|\right)$, and reads up to a constant,
\beg
\rho_2(k,\lam)
&=&
3^{7/6}\Ga(2/3)\left(2\lambda\pi^5\right)^{1/6}
\left[\Bi(0)\Ai\left(-\left(\frac{\pi}{2\lam}\right)^{1/3}|k|\right)
-\Ai(0)\Bi\left(-\left(\frac{\pi}{2\lam}\right)^{1/3}|k|\right)\right]
\nonumber\\
&=&
\pi\,\sqrt{|k|}\,
J_{1/3}\left(\frac{2}{3}\left(\frac{\pi}{2\lam}\right)^{1/2}|k|^{3/2}\right),
\label{airyexpresforro2}
\enq
where we have made use of $\Ai(0)=\frac{3^{-2/3}}{\Ga(2/3)}$,
$\Bi(0)=\frac{3^{-1/6}}{\Ga(2/3)}$ and the relations between
$\Ai(-x),\;\Bi(-x)$ and $J_{\pm 1/3}\left(\frac{2}{3}x^{3/2}\right)$ to
arrive at Eq. (\ref{airyexpresforro2}). Note that $\rho_2(k,\lam)$ vanishes at
$k=0$ and is even in $k.$ We now use the jump discontinuity
of $\frac{d}{dk}\rho_2$ across $k=0$ to determine ${\rm cst}(\lam)$:
$$\cstl = \pk \left[ \,\ak \, \right]_{k=-0}^{k=+0} = \c.$$
Transforming back to $\rho(x,\lambda)$ and with the help of
Eqs. (\ref{recoverro}) and (\ref{expressforrho}) we get
\begin{eqnarray}
\rho(x,\lambda)
&=&
\sqrt{x}\,\int_{0}^{\infty}dk\,\cos(k\,\sqrt{x}) \; \ak + \frac{\c}{\sqrt{x}}.
\label{solutionforthemodesofdiffusionintext}
\end{eqnarray}
Therefore the ``time''
dependent density reads,
\beg
\ro(x,t)=\int_{0}^{\infty}d\lam C(\lam){\rm e}^{-\lam t}\rho(x,\lam),
\enq
where $C(\lam)$ is determined by an initial condition.\\
Eqs. (42) and (43) will now be used to find the dynamical density-density
correlation function.
It is clear that $\cor(x,y,t)$ and
\beg
\icor(x,y,t):=\int_{0}^{y}dz\;\Co(x,z,t)
\enq
satisfy the same diffusion equation as $\rho(x,t)$.
Hence,
\beg
\icor(x,y,t)=\int_{0}^{\infty}d\lam\; C(\lam,y){\rm e}^{-\lam t}
\,\rho(x,\lambda),
\enq
and
\beq
\IIc(x,y,t)& := &
\int_{0}^{x}dz\;\Ic(z,y,t)\\
&=&
\pi\int_{0}^{\infty}d\lam \;C(\lam,y){\rm e}^{-\lam t}
\int_{0}^{\infty}dk\,\sqrt{k}\,\sin(k{\sqrt x})\,\Fl,
\label{timedependenceofintinart}
\enq
where we have used formula (\ref{appsoltodiffresult1})
in appendix \ref{appchechsol} to arrive at the last equation.
$C(\lam,y)$ is determined by the ``initial condition'',
$\IIc(x,y,t=0)=\frac{1}{\bt\pi^2}\ln\left[\frac{|{\sqrt x}-{\sqrt y}|}
{{\sqrt x}+{\sqrt y}}\right]$. A derivation of this formula
can be found in appendix \ref{chvarianceofalinearstatistic}.
To find $C(\lam,y)$ we first perform a Fourier sine transform in the variable
${\sqrt x}$;
$\int_{0}^{\infty}d{\sqrt x}\sin(p{\sqrt x})\cdots,$ on
Eq. (\ref{timedependenceofintinart}) at $t=0$ to get an integral
equation satisfied by $C(\lam,y)$\footnote{We use the transform
$\int_{0}^{\infty}dx\ln\bigl|\frac{x+a}{x-a}\bigr|\sin(px)=\frac{\pi}{p}
\sin(pa).$}
\beg
-\frac{1}{\bt\pi\,p}\sin(p{\sqrt y})=\frac{\pi^2}{2}\int_{0}^{\infty}
d\lam C(\lam,y){\sqrt p}J_{1/3}\left(\frac{2}{3}{\sqrt {\frac{\pi}{2\lam}}}
p^{3/2}\right).
\enq
This can be rewritten as
\beg
-\frac{2\sin(p{\sqrt y})}{\bt\pi^4p^{3/2}}=
{\cal H}_{1/3}\left(\frac{C\left(\frac{x}{2u^2},y\right)}{u^4},\frac{2}{3}
p^{3/2}\right),
\enq
where ${\cal H}_{\nu}(f(u),\xi)={\bar f}_{\nu}(\xi)
:=\int_{0}^{\infty}du f(u)uJ_{\nu}(u\xi),$
denotes the Hankel transform. Using the Hankel inversion theorem
[$f(x)=\int_{0}^{\infty}d\xi{\bar f}_{\nu}(\xi)\xi J_{\nu}(\xi u)$,
see\cite{Sneddon}], we find,
\beg
\frac{C\left(\frac{\pi}{2u^2},y\right)}{u^4}
=-\frac{4}{3\bt\pi^4}\int_{0}^{\infty}dp\sin(p{\sqrt y}){\sqrt p}
J_{1/3}\left(\frac{2}{3}p^{3/2}u\right).
\enq
Therefore the ``time'' dependence of the twice integrated correlation
function
is displayed as a Laplace transform with the spectral parameter
$\lam:=\frac{\pi}{2u^2}$,
\beg
\IIc(x,y,t)=-\frac{1}{3\bt\pi^2}\int_{0}^{\infty}\frac{d\lam}{\lam^2}
{\rm e}^{-\lam t}\,F(x,\lam)F(y,\lam),
\enq
where $F(x,\lam):=\int_{0}^{x}dz\ro(z,\lam)$ and $\ro(z,\lam)$ given by
Eqs. (27) and (42).
Thus the variance of an arbitrary linear statistics,
\beg
{\cal Q}(t)
&:=&
\sum_{a}{\cal Q}(x_a(t))
=\int_{0}^{\infty}dx\sum_{a}\delta(x-x_a(t)){\cal Q}(x),
\enq
is
\beq
{\rm Var}[{\cal Q},t]
&:=&
<\cq(t)\cq(0)>_\eq-<\cq(t)>_\eq<\cq(0)>_\eq
\enq
\beq
= \int_{0}^{\infty}dx\int_{0}^{\infty}dy
\cq(x)\cq(y)\Co(x,y,t)
=\int_{0}^{\infty}dx\int_{0}^{\infty}dy\cq^{\prime}(x)\cq^{\prime}(y)
\IIc(x,y,t).
\enq
This is the main result of our paper.\\
\noindent
{\bf  4 Variance of} $\frac{1}{1+x}${\bf .}
\par\noindent
As an example we take $\cq(x)=\frac{1}{1+x},$ which is the Landauer formula
for the conductance. A simple calculation gives
\beg
\vr\left[\frac{1}{1+x},t\right]=\frac{\pi}{3\bt}\int_{0}^{\infty}
\frac{d\lam}{\lam^2}{\rm e}^{-\lam t}h^2(\lam),
\enq
with
$$
h(\lam):=\frac{1}{\pi^2}\int_{0}^{\infty}\frac{dx}{(1+x)^2}F(x,\lam)
$$
\beg
=\frac{1}{2}\int_{0}^{\infty}dkk^{3/2}{\rm e}^{-k}\Fl
=\left(\frac{2\lam}{\pi}\right)^{5/6}
\sum_{n=0}^{\infty}A_n\left(\frac{18\lam}{\pi}\right)^{n/3},
\enq
where
\beg
A_n:=\frac{(-1)^n}{n!}\frac{\Ga(1+n/3)}{\Ga\left(-(n-1)/3\right)}.
\enq
Eqs. (56) and (57) can be found in\cite{Prudnikov}.
Now when $n\equiv 1\;{\rm mod}\;3$, $A_n$ is taken to be zero. The remaining
terms can be grouped into two sums with $n\equiv 0\;{\rm mod}\;3$ and
$n\equiv 2\;{\rm mod}\;3$ respectively. An application of the ratio test
to these series shows that both converge absolutely and uniformly for all
$\lam.$ We deduce that (with $\bt=2$),
\beg
\frac{\pi}{6}\;\frac{h^2(\lam)}{\lam}
=\lam^{2/3}\sum_{n=0}^{\infty}B_n\lam^{\frac{n}{3}},
\enq
converges uniformly in the $\lam^{1/3}$ plane. Using this representation
and after performing the Laplace transform, we get,
\beg
\vr\left[\frac{1}{1+x},t\right]=\frac{1}{t^{2/3}}
\sum_{n=0}^{\infty}B_n\frac{\Ga\left(\frac{n+2}{3}\right)}
{t^{\frac{n}{3}}}
=\frac{C_{2/3}}{t^{2/3}}+\frac{C_{4/3}}{t^{4/3}}+\frac{C_{5/3}}{t^{5/3}}+
\frac{C_{2}}{t^2}+\frac{C_{7/3}}{t^{7/3}}+\cdots,
\enq
where
$$C_{2/3}=\frac{\left(\frac{3}{2\pi^2}\right)^{1/3}
\Ga(2/3)}{2\Ga^2(1/3)}\approx 0.050346,$$
$$C_{4/3}=\frac{3\left(\frac{3}{2\pi^2}\right)^{2/3}
\Ga(5/3)\Ga(4/3)}{\Ga(-1/3)\Ga(1/3)}\approx
-0.0632871,$$
$$
C_{5/3}=-\frac{3\left(\frac{3}{2\pi^5}\right)^{1/3}\Ga(5/3)}{\Ga(1/3)\Ga(-2/3)}
\approx 0.042735,$$
$$
C_{2}=\frac{27\Ga^2(5/3)}{4\pi^2\Ga^2(-1/3)}\approx 0.0337737, $$
$$C_{7/3}=-\Ga(7/3)
\left[\frac{9\left(\frac{9}{4\pi^7}\right)^{1/3}\Ga(5/3)}{\Ga(-1/3)\Ga(-2/3)}-
\frac{9\left(\frac{9}{4\pi^7}\right)^{1/3}\Ga(8/3)}{10\Ga(-4/3)\Ga(1/3)}\right]
\approx -0.0716263.$$
This completes the determination of the long ``time'' behaviour
of $\vr[1/(1+x),t]$. According to the Watson Lemma\cite{Watson},
Eq. (59) is an asymptotic expansion. The fact the Brownian motion ensemble
tends to the stationary ensemble  with exponential speed as soon as
$\ta D/\ga L\gg 1,$ enables us to conclude that
$\lim_{t\to \infty}\vr[\cq,t]=0.$
At $t=0,$ we simply go back to Eq.(55)
and make use of the completeness theorem of the Hankel transform to
show that $\vr[1/(1+x),0]=\frac{1}{8\bt},$ a result obtained previously
\cite{Beenakeruniversal,Basor,linstat}.
Other examples of $\cq$ can be found in\cite{Beenakeruniversal}.

\vfill\eject
\appendix
\section{Check of solution}
\label{appchechsol}
In this appendix we are going to show that
\begin{eqnarray}
\rho(x,\lambda)
& = &
\frac{\c}{\sqrt{x}} + \int_{0}^{\infty}\, \sqrt{x}\,\cos(k\, \sqrt{x}) \, \ai
\end{eqnarray}
is a solution to the integro-differential equation
\begin{eqnarray}
\lambda\,\rho(x,\lambda)
&=&
\px \frac{1}{\sqrt{x}} \int_{0}^{\infty} \frac{\rho(y,\lambda)}{x- y } dy.
\label{appfokkerplankhalvakse}
\end{eqnarray}
To obtain better convergence in the manipulations
we will instead show $\rho(x,\lambda)$ fulfills the equation
obtained by integrating Eq.(\ref{appfokkerplankhalvakse}) from
$0$ to $z$. From the boundary condition Eq.(\ref{fluxboundarycondition})
we see that the lower boundary term on the right hand side to vanish.
Using this, Eq.(\ref{appfokkerplankhalvakse}) becomes
\begin{eqnarray}
\lambda\,\int_{0}^{z}\,dx \,\rho(x,\lambda)
&=&
\frac{1}{\sqrt{z}} \int_{0}^{\infty} \frac{\rho(y,\lambda)}{z - y } dy.
\label{appintegralfokkerplankhalvakse}
\end{eqnarray}
This equation is a stronger statement than
Eq. (\ref{appfokkerplankhalvakse}) since any constant will vanish when
Eq. (\ref{appintegralfokkerplankhalvakse}) is differentiated to give
Eq. (\ref{appfokkerplankhalvakse}).
We start by evaluating the right hand side of Eq.(62),
\begin{eqnarray}
\mbox{R.H.S}
&=&
\frac{1}{\sqrt{z}} \int_{0}^{\infty} \frac{\rho(y,\lambda)}{z - y } dy\\
&=&
\frac{1}{\sqrt{z}} \int_{0}^{\infty} \frac{\c}{z - y }\frac{dy}{\sqrt{y}}
+
\frac{1}{\sqrt{z}} \int_{0}^{\infty} \,dy \int_{0}^{\infty}\,
\frac{\sqrt{y}\,\cos(k\, \sqrt{y})}{z-y} \, \ai
\nonumber\\
&=&
\frac{1}{\sqrt{z}} \int_{0}^{\infty} \,dy \int_{0}^{\infty}\,
\frac{\sqrt{y}\,\cos(k\,\sqrt{ y})}{z-y} \, \ai.
\end{eqnarray}
Using the identity,
$$
\int_{0}^{\infty}\, \frac{\sqrt{y}\,\cos(k\, y)}{z-y} \,dy
=
2\pi\,\delta(k) + \pi\,\sqrt{z} \sin(k\,\sqrt{z}),
$$ in Eq. (64) gives
\begin{eqnarray}
\frac{1}{\sqrt{z}} \int_{0}^{\infty} \frac{\rho(y,\lambda)}{z - y } dy
&=&
\pi \,\int_{0}^{\infty}\,\sin(k\,\sqrt{z}) \,\ai.
\end{eqnarray}
\noindent
We now turn to the left hand side of Eq. (\ref{appintegralfokkerplankhalvakse})
\begin{eqnarray}
\mbox{L.H.S}&=&\lambda\,\int_{0}^{z}\,dx \,\rho(x,\lambda)\\
&=&
\lambda\int_0^{z}\frac{\c\,dx}{\sqrt{x}}
+
\lambda\int_0^{z}dx\,\int_{0}^{\infty}\, \sqrt{x}\,\cos(k\, \sqrt{x}) \, \ai
\nonumber\\
&=&
2\,\lambda\,\c\,\sqrt{z}
+
2\,\lambda\,\int_{0}^{\infty}\,\int_0^{\sqrt{z}}\,dt\,t^2\,\cos(k\,t)\,\ai.
\end{eqnarray}
The integral with respect to $t=\sqrt{x}$ is
\begin{eqnarray}
\int_0^{\sqrt{z}}\,dt\,t^2\,\cos(k\,t)
&=&
-\pkt \left\{\, \frac{\sin(k\,\sqrt{z})}{k}\right\}.
\end{eqnarray}
The left hand side is then
\begin{eqnarray}
\mbox{L.H.S}
&=&
2\,\lambda\,\c\,\sqrt{z}
-
2\,\lambda\int_{0}^{\infty}\,
\pkt \left\{\, \frac{\sin(k\,\sqrt{z})}{k}\right\}\,\ai
\nonumber\\
&=&
2\,\lambda\,\c\,\sqrt{z}
-2\,\lambda\,\left[\pk
\left\{\,
\frac{\sin(k\,\sqrt{z})}{k}\right\}\,\ak\right]_{0}^{\infty}\nonumber\\
&& + 2\,\lambda\int_{0}^{\infty}\,
\pk \left\{\,
\frac{\sin(k\,\sqrt{z})}{k}\right\}\,\pk\left\{\,\ak\,\right\}\,dk.
\end{eqnarray}
The boundary terms from the partial integration are zero,
since
$$
\ak\sim {\rm const}\,k^{-\frac{1}{4}}
\cos({\rm const}\,k^{\frac{3}{2}}-{\rm phase}),\;k\rightarrow \infty
$$
\begin{equation}
\;\;\;\;\;\;\;\;=0,\;\;\;k=0.
\end{equation}
Now,
\begin{eqnarray}
\mbox{L.H.S}
&=&
2\,\lambda\,\c\,\sqrt{z}
+
2\,\lambda\,\left[\, \frac{\sin(k\,\sqrt{z})}{k}\,
\pk\left\{\,\ak\,\right\}\right]_{0}^{\infty}
\nonumber\\
&&
-2\,\lambda\int_{0}^{\infty}\,\frac{\sin(k\,\sqrt{z})}{k}\,
\pkt\left\{\,\ak\,\right\}\,dk.
\end{eqnarray}
The boundary term at infinity vanishes. The boundary term at the origin
is
\begin{eqnarray}
- 2\,\lambda\,\sqrt{z}\,\left.\pk\left\{\,\ak\,\right\}\right|_{k=0}
&=&
- 2\, \lambda\,\c\,\sqrt{z}.
\end{eqnarray}
By differentiation,
\begin{eqnarray}
\frac{1}{k}\,
\pkt\left\{\,\ak\,\right\}
&=&
- \frac{\pi}{2\,\lambda}\, \ak.
\end{eqnarray}
Gathering the pieces we have
\begin{eqnarray}
\lambda\,\int_{0}^{z}\,dx \,\rho(x,\lambda)
&=&
\pi \,\int_{0}^{\infty}\,\sin(k\,\sqrt{z}) \,\ai
\label{appsoltodiffresult1}\\
\lambda\,\int_{0}^{z}\,dx \,\rho(x,\lambda)
&=&
\frac{1}{\sqrt{z}} \int_{0}^{\infty} \frac{\rho(y,\lambda)}{z - y } dy
\label{appchecksoltodifresult}
\end{eqnarray}
Here we make two remarks on the above derivations:\\
1. Since the number of particles in the interval
$(0,z)$ goes to zero, when $z$ tends to zero, the last equation
shows that the zero flux condition at the origin is fulfilled.\\
2. This is concern with the $z\rightarrow \infty$ limit
in Eq. (\ref{appsoltodiffresult1}).
The integrand on the right hand side is the product of two strongly
oscillating functions,
which oscillate out of phase. Since none of them are
absolutely integrable, a Riemann-Lebesque lemma can not be used to conclude
that the integral tends to zero. We may appeal to
the theory of generalized functions that it is 0.
The R. H. S. of Eq. (74) can be re-written up to irrelevant constants as
$$ {\cal J}(a)=\int_{0}^{\infty}dy\;\sin(ya){\sqrt y}J_{1/3}(by^{3/2}),$$
where $a:={\sqrt z}$ and
$b:=\frac{2}{3}\left(\frac{\pi}{2\lam}\right)^{1/2}.$ We wish to determine
$\cj(a)$ in the limit $a\rightarrow \infty,$ with $b (>0)$ fixed.
Consider instead,
$$\cj_{\mu}(a)=\int_{0}^{\infty}dy \frac{\sin(ya)}{y}y^{1-\mu}
J_{1/3}(by^{3/2}),\;\; -\frac{5}{4}<\Re\mu<\frac{5}{2}.$$
The integral we need can be defined as the analytic continuation
of $\cj_{\mu}(a)$ to $\mu=-1/2.$ Now, since
$\lim_{a\to\infty}\frac{\sin(ax)}{x}=\delta(x),$ we conclude by
integrating over the $\delta$ function that
$\lim_{a\to\infty}\cj_{\mu}(a)=0.$ Hence $\cj(\infty)=0.$
Physically this means that the total number
of particles in each lambda mode is zero. The zero mode, of course,
contains particles.
\vfill\eject
\section{The equal time variance of a linear statistic.}
\label{chvarianceofalinearstatistic}
The equal time variance of a linear statistic $\cq$ is
\begin{eqnarray}
\Var(\cq,0) &=&
\int_{0}^\infty dx \int_{0}^\infty dy \,\cq(x)\,\cq(y) \, \cor(x,y,0),
\label{equaltimevarianceexpressedwithcor}
\end{eqnarray}
a result first derived in \cite{Beenakeruniversal} by the method of
functional derivative. The claim in \cite{Beenakeruniversal} is that the
method is valid for all potentials, however,
this is flawed by the use of the $\n=\infty$ density, which may not exist
for {\it all} potentials. For the
sake of completeness, we present in this
appendix a small modification of the arguments given in
\cite{Beenakeruniversal} and show that the result has a more general
validity. Instead of focusing on
Eq. (\ref{equaltimevarianceexpressedwithcor}),
we prefer to work with the partially integrated expressions
\begin{eqnarray}
\Var(\cq,t) &=& -
\int_{0}^\infty dx \int_{0}^{\infty} dy \,\cq(x)\,\cq^{\prime}(y) \,
\icor(x,y,t),\;\;\;
\label{timevarianceexpressedwithicor}
\end{eqnarray}
and
\begin{eqnarray}
\Var(\cq,t) &=&
\int_{0}^\infty dx \int_{0}^{\infty} dy \,\cq^{\prime}(x)\,\cq^{\prime}(y) \,
\iicor(x,y,t),\;\;\;\footnotemark
\label{timevarianceexpressedwithiicor}
\end{eqnarray}
\footnotetext{To avoid the boundary terms it has been
assumed that the linear statistic fulfills
$\left.\sqrt{x}\,\cq(x)\right|_{x=0}=0$. The linear
statistic corresponding to the conductance $\frac{1}{1+x}$
satisfies this criterion.}
where
\begin{eqnarray}
\icor(x,y,t) &\equiv& \int_{0}^{y} dz \,
\cor(x,z,t),\\
\iicor(x,y,t) &\equiv& \int_{0}^{x} dz \,
\icor(z,y,t) = \int_{0}^{x} dz_1 \int_{0}^{y} dz_2 \,
\cor(z_1,z_2,t).
\end{eqnarray}
We will start by finding $\icor(x,y,0)$.
By definition
$\sigma_\gas (x) = \sum_{n=1}^{\infty} \delta(x-x_{n}^{0})$
and
\begin{eqnarray}
<\,\sigma_\gas (x) \,>_\eq &=& \frac{\int_0^{\infty} dx_{1}^{0}\ldots
\int_0^{\infty} dx_{\n}^{0}\, \sigma_\gas (x)\,
e^{-W[u]}}
{\int_0^{\infty} dx_{1}^{0}\ldots\int_0^{\infty} dx_{\n}^{0}\,e^{-W[u]}}.
\end{eqnarray}
The dependence of the energy, $W$, on the coordinates
$x_{1}^{0},\,x_{2}^{0},\ldots,x_{\n}^{0}$ has been suppressed.
The external potential $u(x)$ is assumed to be bounded at the origin
which is the physically interesting situation. Now write
$u(x) = -\int_0^x f(z) \,dz + u(0)$, where $f(x)$ is the force.
The constant $u(0)$ can be set to zero as this can always be
accomplished by a redefinition of the zero point energy. The energy
is now
\begin{equation}
W[u]=W\left[-\int_0^x f(z)dz\right]=-\sum_{n=1}^{\n}\int_0^\infty f(z)\,
\heavi(x_n^{0}-z) \,dz
-\,\sum_{i <j} \ln|x_{j}^{0} - x_{i}^{0}|,
\end{equation}
where $\heavi$ is the Heaviside step function.
The functional derivative of $W[-\int_0^x f(z)\,dz]$ with respect to
$f$ is
\begin{eqnarray}
\frac{\delta W[-\int_0^x f(z)\,dz]}{\delta f(y)} &=&
- \sum_{n=1}^{\infty} \heavi(x_n^{0}-y)
=
- \int_0^{y}\sigma_\gas (x,0)\,dz.
\end{eqnarray}
The functional derivative of $\sigma_\eq(x) = <\,\sigma_\gas(x,0) \,>_\eq$
with respect to $f$ is recognized as $\icor(x,y,0)$,
\begin{eqnarray}
\icor(x,y,0) &=& \frac{\delta \,\sigma_\eq(x) }{\bt\delta f(y)}.
\end{eqnarray}
\noindent
The density $\sigma_\eq(x)$ is approximated
by
\beq
\sigma(x,\n;f) &=&
- \frac{1}{\pi^2}\,\sqrt{\frac{b(\n,f) - x}{x}}
\int_0^{b(\n,f)}
\frac{dz}{z-x}\,\sqrt{\frac{z}{b(\n,f) - z}}
\,\, f(z),
\enq
where $b(\n,f)$ is the upper limit of support of $\sigma(x,N;f)$.
$\frac{\delta \,\sigma(x,\n;f) }{\delta f(y)}$ is computed as the linear term
in $\epsilon$, when $\epsilon\,\delta(x-y)$ is added to
the force $f(x)$.
We have
\begin{eqnarray}
\sigma(x,\n;f +\epsilon\,\delta(x-y))
&=&
- \frac{1}{\pi^2}\,\sqrt{\frac{b - x}{x}}
\int_0^{b}
\frac{dz}{z-x}\,\sqrt{\frac{z}{b - z}}
\,\,[\;f(z) + \epsilon\,\delta(z-y)\;]\\
&=&  \epsilon\;\frac{1}{\pi^2}\,\sqrt{\frac{y}{x}}\,\frac{1}{x-y}
\,\;\;\sqrt{\frac{b - x}{b - y}}
+ \sigma(x,\n-\eta;f),
\end{eqnarray}
where $\eta = \eta(\epsilon,y,\n,f) =\epsilon\;\frac{1}{\pi^2}\, \int_0^{b}
\sqrt{\frac{y}{x}}\,\frac{1}{x-y}\,
\sqrt{\frac{b - x}{b - y}}\,dx = -\epsilon \, \frac{1}{\pi}\,\frac{y}{b-y} $
is the
number of particles associated with the force $\epsilon\,\delta(x-y)$
and $b:=b(\n,f +\epsilon\,\delta(x-y)) = b(\n-\eta,f)$
is the upper limit of support of $\sigma(x;\n,f +\epsilon\,\delta(x-y))$
and $\sigma(x,\n-\eta;f)\,$\footnote{
They have the same upper limit, because the part of
$\sigma(x,\n;f +\epsilon\,\delta(x-y))$ corresponding to $f$:
$- \frac{1}{\pi^2}\,\sqrt{\frac{b - x}{x}}
\int_0^{b}
\frac{dz}{z-x}\,\sqrt{\frac{z}{b - z}}\, f(z)$
is a solution, with $\n-\eta$ particles, to the problem with
the  external force $f$, obeying the right boundary conditions. In other
words  $- \frac{1}{\pi^2}\,\sqrt{\frac{b - x}{x}}
\int_0^{b}
\frac{dz}{z-x}\,\sqrt{\frac{z}{b - z}}\, f(z) = \sigma(x,\n-\eta;f)$.}.
The functional derivative is
\begin{eqnarray}
\frac{\delta \,\sigma(x,\n) }{\bt\delta f(y)}
&=&
\frac{1}{\bt\pi^2}\,\sqrt{\frac{y}{x}}\,\frac{1}{x-y}\;\;\frac{b-x}{b-y}
\,+\, \frac{1}{\bt\pi}\,\frac{y}{b-y} \;\;\;
\frac{\partial \,\sigma(x,\n) }{\partial \n}.
\end{eqnarray}
In the limit $\n \rightarrow \infty$,
$\sqrt{\frac{b - x}{b - y}} =1$, $\eta$ and
$\frac{\partial \,\sigma(x,\n) }{\partial \n}$ tend to zero.
The extra density, $\Delta\sigma(x),$ when $\epsilon$ particles is added
to the system, experiences no force in the interval
$(0,b)$. Therefore $\Delta \sigma(x)$ is less than
the corresponding density with
$\epsilon$ particles in the box $(0,b)$:
$\frac{1}{\pi}\frac{\epsilon}{\sqrt{x(b-x)}},\;\;x\in(0,b).$
Outside $(0,b)$, $\Delta \sigma(x)$ tends monotonically to zero. We conclude,
in this limit, the functional derivative is reduced to
\begin{eqnarray}
\frac{\delta \,\sigma(x,\n) }{\bt\delta f(y)}
&=&
\frac{1}{\bt\pi^2}\,\sqrt{\frac{y}{x}}\,\frac{1}{x-y}\\
\icor(x,y,0)
&=&
\frac{1}{\bt\pi^2}\,\px\,\ln\left|\frac{\sqrt{x} - \sqrt{y}}
{\sqrt{x} + \sqrt{y}}\right|.
\label{expresforequaltimeicor}
\end{eqnarray}
{}From this it is seen that
\begin{eqnarray}
\iicor(x,y,0)
&=&
\frac{1}{\bt\pi^2}\,\ln\left|\frac{\sqrt{x} - \sqrt{y}}
{\sqrt{x} + \sqrt{y}}\right|
\label{expresforequaltimeiicor}
\end{eqnarray}
and
\begin{eqnarray}
\Var(\cq,0)
&=&
\frac{1}{\bt\pi^2}\,\int_{0}^\infty dx \int_{0}^{\infty} dy
\,\cq^{\prime}(x)\,\cq^{\prime}(y) \,
\ln\left|\frac{\sqrt{x} - \sqrt{y}}
{\sqrt{x} + \sqrt{y}}\right|.
\end{eqnarray}
Thus at equal time, variances are independent of the potential $u$.
\vfill\eject


\begin{thebibliography}{99}
\bibitem{GFP} The asymptotic gap formation probability reads,
$$E_{\bt}(0,s)\sim G(1+\alpha)\frac{{\rm e}^{-\frac{s}{2\bt}+
\frac{2\alpha{\sqrt s}}{\bt}}}{s^{\frac{\alpha(2-\bt)}{4\bt}+
\frac{\alpha^2}{2\bt}}}\left[1+{\rm O}\left(\frac{1}{\sqrt s}
\right)\right],$$
where $G(.)$ is the Barnes $G-$function.
\end{thebibliography}
\end{document}